\documentclass[a4paper,showkeys,floatfix,aps,prl,reprint,groupedaddress]{revtex4-1}
\usepackage{mathtools}
\usepackage{amsmath,amssymb}
\usepackage{graphics,graphicx}
\usepackage{dcolumn,bm}
\usepackage{psfrag}
\usepackage{color}
\begin{document}

\title{Inequalities of energy release rates in compression of nano-porous materials predict its imminent breakdown}

\author{Diksha}
\affiliation{Department of Physics, SRM University - AP, Andhra Pradesh 522240, India}

\author{Jordi Bar\'{o}}
\affiliation{Departamento de Física de la Materia Condensada
Facultat de Física, University of Barcelona, Spain.}

\author{Soumyajyoti Biswas}
\affiliation{Department of Physics, SRM University - AP, AndhraPradesh 522240, India}

\begin{abstract}
We show that the divergent acoustic energy release rate in a quasi-statically compressed nano-porous material can be used as a precursor to failure in such materials. A quantification of the inequality of the energy release rate using social inequality measure indices help constructing a warning signal for large bursts of energy release. We also verify similar behavior for simulations of viscoelastic fiber bundle models that mimic the strain-hardening dynamics of the samples. The results demonstrate experimental applicability of the precursory signal formulation for any diverging response function near a transition point using social inequality indices. 
\end{abstract}

\maketitle
A robust precursory signal for an imminent large response for a driven disordered system is a long standing problem. In the context of disordered systems, such problems relate to the extreme statistics of fracture -- from the laboratory scale of rock samples to the tectonic scale of earthquakes \cite{earthquake,kagan}. There is a plethora of rule-based (see e.g., \cite{alava,karppinen,graham,saichev}) and machine learning (see e.g., \cite{cubuk,bapst,creep,PhysRevE.106.025003}) forecasting techniques that attempt in addressing this problem with various degrees of successes. What makes the problem difficult, however, is the general lack of number of samples \cite{Zoller,pnas18} and in particular the lack of access to the measuring quantities in the vast majority of the real-life examples. Therefore, even though a particular method might be successful in theory, its implementations in experiments remain challenging. 

A recent method in predicting catastrophic changes in a system near a transition point was developed based on the quantification of the inequalities of a (diverging) response function through measuring indices commonly used in economics viz. the Gini index ($g$) \cite{gini} and the Kolkata index ($k$) \cite{kolkata}. It was shown for a broad class of systems that near any (equilibrium or non-equilibrium) critical point, the scaling form of various response functions undergo drastic simplification when written in terms of these indices and that the crossing point of the two inequality indices indicate a precursory signal of the imminent diverging response (a catastrophic change in some cases) \cite{soumyaditya, PhysRevE.109.044113}. The analytical results were also tested numerically in several systems, including the fiber bundle model of fracture \cite{soumyaditya,PhysRevE.109.044113,PhysRevE.108.014103}. 

In this work, we apply the above mentioned method for the case of diverging energy release rate in compression of nano-porous materials (silica based materials and natural rock). It is known that the elastic energy release rate show divergence prior to large events \cite{Prl_18}. Therefore, the inequality of the release rate can be used to show the precursory signal for each of those large event. We also show the precursory signals in the case of simulation of viscoelastic fiber bundle model \cite{PRE_jordi} that is known to mimic the dynamics of such materials. 

In the cases where the size distribution of a response to external driving (say, avalanches in response to compression loading) is a power law of the form $P(S)=CS^{-\delta}$, it is straightforward to see that for a series of $m$ such events arranged in ascending order, the size of the $r$-th ranked event will follow $S_r\propto(m-r)^{-n}$, where $n=1/(1-\delta)$ and $\delta>1$ \cite{PhysRevE.109.044113}. It is then possible to measure the inequality of these events by constructing what is called a Lorenz curve. The Lorenz function $\mathcal{L}(p,n)$ \cite{lorenz} refers to the ratio of the cumulative event (avalanche) mass of the smallest $p$ fraction of the series and the total event (avalanche) mass of the series. Due to the ordering operation, this would simply be $\mathcal{L}(p,n)=\frac{\int\limits_0^{pm} (m-r)^{-n} dr}{\int\limits_0^{m} (m-r)^{-n} dr}$. Now, the entire series already contains the largest event ($S_m$ after ordering) in the series, which is what we are aiming at predicting. Therefore, we need to consider the first $b$ fraction of the series (and then vary $b$). In that case, the Lorenz function reads $\mathcal{L}(p,b,n)=\frac{1-(1-pb)^{1-n}}{1-(1-b)^{1-n}}$. 

\begin{figure*}
    \includegraphics[width=16cm]{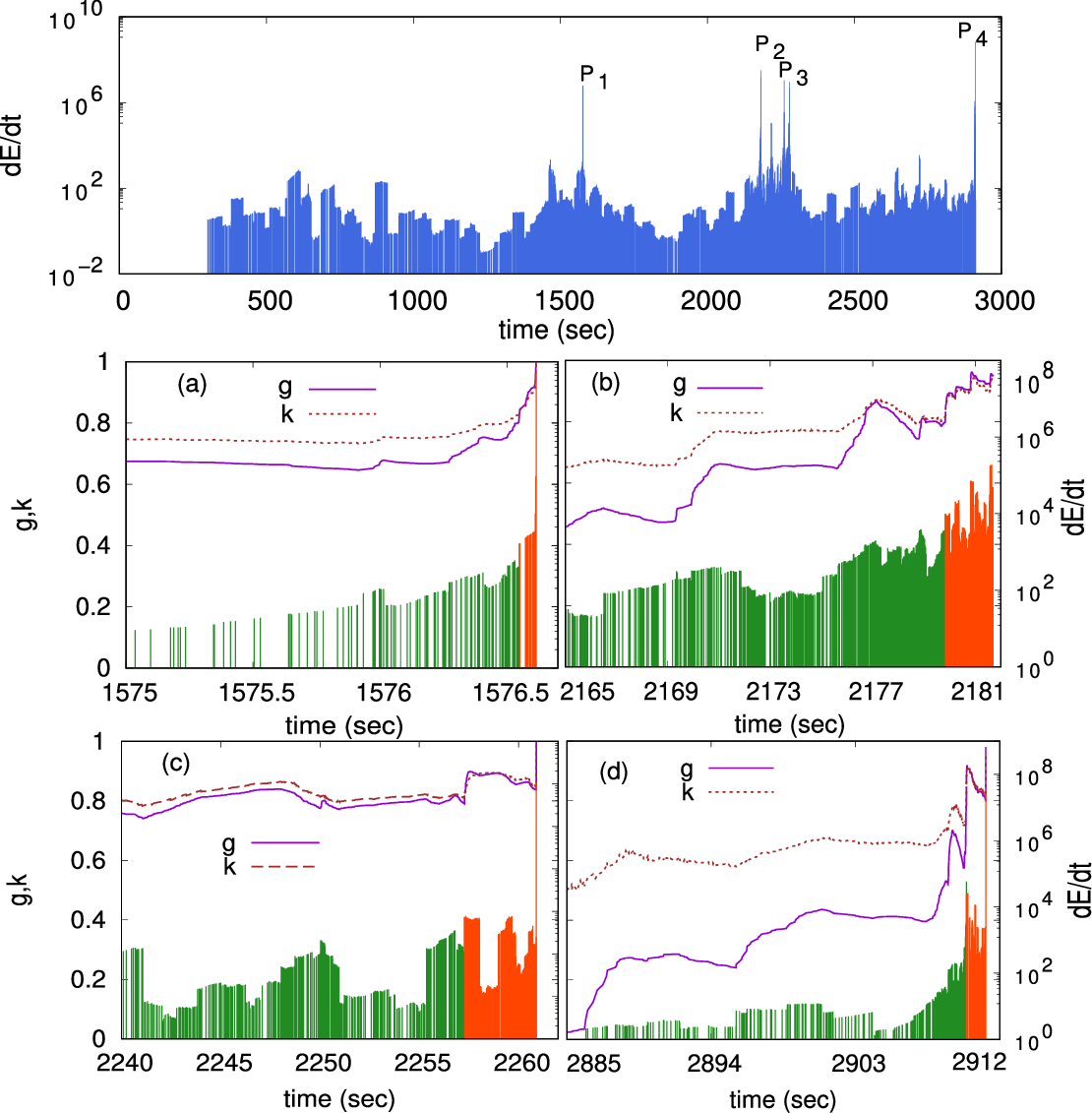}
    \caption{The Energy release rate of the acoustic emission is shown along with time using a moving window of 50 for the natural red sandstone sample (SR2). In Figures (a-d), the temporal evolution of $g$ and $k$ is shown for each peak. We noted that for each peak ($P_{1}-P_{4}$), the crossing of $g$ and $k$ occurs before the occurrence of the significant peak.we have indicated all events following the crossing with red color and all events prior to the crossing with green color. This intersection serves as a precursor signal preceding failure in all cases.}
    \label{sr2}
\end{figure*}

The definition of Lorenz function helps quantifying inequality in the series of values. Note that in the extreme case of all events being exactly of the same size, the Lorenz function by its definition would be a diagonal straight line (the equality line) between $p=0$ ($\mathcal{L}=0$) to $p=1$ ($\mathcal{L}=1$). For unequal events, such as the case above, the Lorenz function is still bounded between those two values, but increases non-linearly (a concave function), monotonically and continuously in between. Therefore, the area between the equality line and the Lorenz curve is a measure of inequality of the particular series considered. The Gini index $g$ is such a measure (normalized by the total area under the equality line, 1/2): $g(b,n)=1-2\int\limits_0^1\mathcal{L}(p,b,n)dp$. The other inequality index we are interested in is the Kolkata index ($k$), which is defined from the equation $1-k=\mathcal{L}(k)$ and it signifies that $1-k$ fraction of the largest avalanches accounts for $k$ fraction of the total avalanche mass (or damage). This is a generalisation of the Pareto's 80-20 law \cite{Pareto}. The expressions for the Gini ($g$) and Kolkata ($k$) indices, in the particular case of a power-law distributed set of events, would be 
\begin{eqnarray}
     g(b,n) &=&1-\frac{2}{1-(1-b)^{1-n}}\left[1+\frac{(1-b)^{2-n}-1}{(2-n)b}\right], \nonumber \\
      1-k(n,b) &=& \frac{1-(1-k(n,b)b)^{1-n}}{1-(1-b)^{1-n}}.
      \label{gk_expressions}
\end{eqnarray}
The value of $k$ requires numerical evaluation, and the expressions do not work for $n=1$ and $n=2$ \cite{soumyaditya}. 

The functions $g$ and $k$ cross each other as $b$ is increased towards $b=1$, as long as $\delta<2$ (recall $n=1/(1-\delta)$). At this point we have to make the approximation that a large event is preceded by a series of events of growing sizes (on average). This is clear when we consider a diverging response function (say, susceptibility in Ising model), for example. In that case, as the driving field (say, temperature) is tuned gradually towards the critical point, the function grows on average, and $g$ and $k$ cross prior to the critical point is reached. 

\begin{figure}
    \includegraphics[width=10cm]{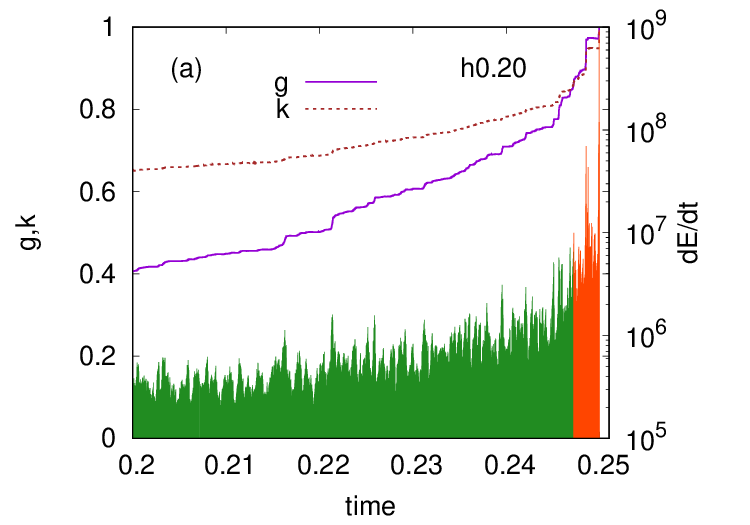}
     \includegraphics[width=10cm]{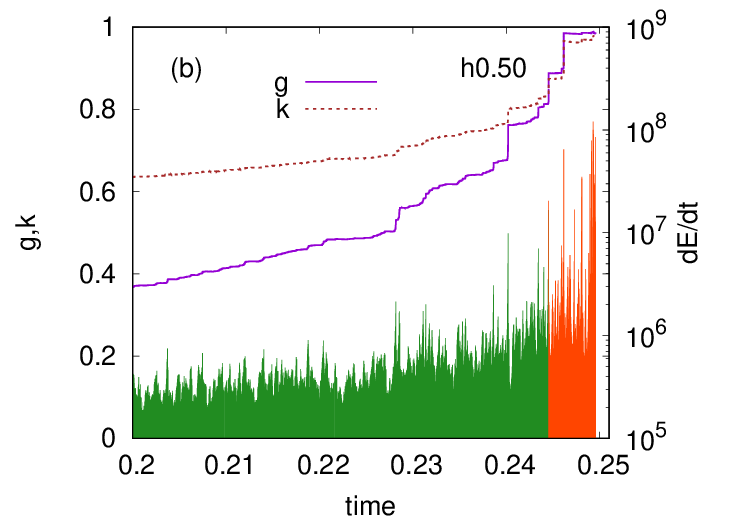}
      \includegraphics[width=10cm]{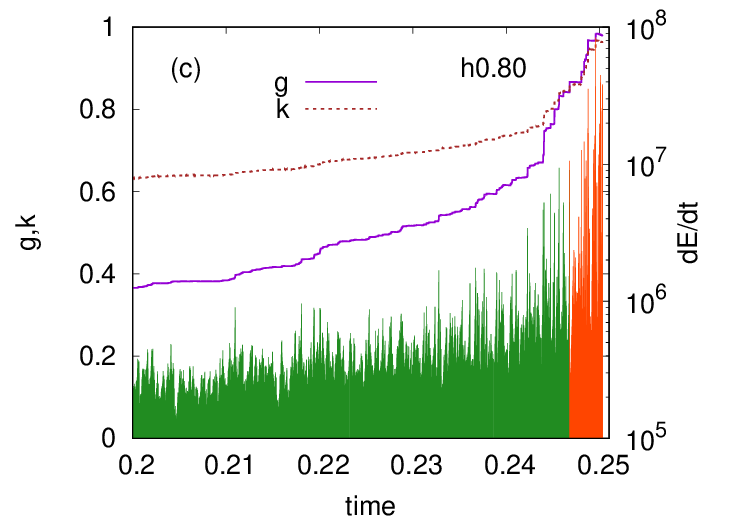}
    \caption{The time series of energy release rate (in logarithmic scale) is shown along with the time variations of $g$ and $k$ for the RFBM for different value of $h$.}
    \label{model}
\end{figure}

In the (fiber bundle) model of fracture \cite{pierce}, avalanche series is the response due to small increases in the external load. In mean-field version, it can be shown that while the sizes leading upto the large catastrophic event are not strictly monotonic, they are on average a growing (diverging) quantity. It was also shown, for non mean-field interface propagation models that the event leading upto the depinning event are growing in size on average. In these cases, therefore, the parameter $b$ serves as a proxy to the applied load (assuming load is linearly increasing with time) or time. The crossing of $g$ and $k$ can then serve as a precursor to imminent failure (or large event) \cite{PhysRevE.109.044113}. 

\begin{figure*}
    \includegraphics[width=7cm]{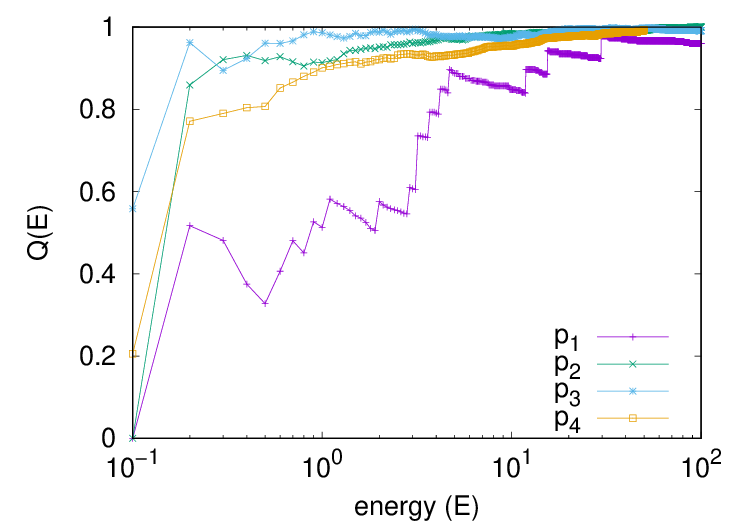}
    \includegraphics[width=7cm]{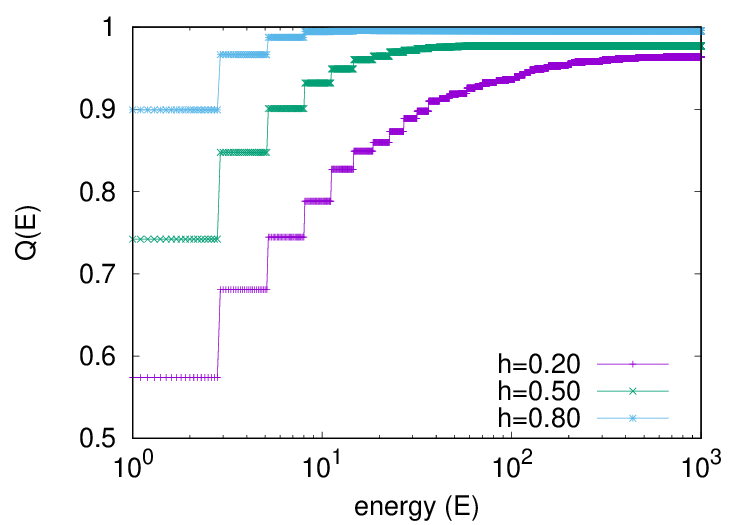}
    \caption{The cumulative fraction of post crossing events for experiment and simulation are shown for each peak as shown in Fig.\ref{sr2} and Fig.\ref{model}. The plots shows the correlation between large size events following the crossing.}
    \label{Q_measures}
\end{figure*}

However, it is not guaranteed that in all cases of fracture and breakdown the average avalanche sizes will grow prior to a major event. The particular case of the nano-porus rocks that we study here, is indeed a case where the avalanche (acoustic emission) sizes do not grow prior to a large event (drop in height of the sample). But it was shown previously that the rate of events grow prior to a large event \cite{Prl_18}. This means that the signature divergence near a critical point could be seen in the acoustic energy emission rate (within a suitable time window) rather than the acoustic emission energies themselves. The series show a build-up prior to large events that are clearly identifiable with the major peaks in the acoustic emission data. 
 
As mentioned before, The experimental data corresponds to the acoustic emission (a.e.) events recorded from the mechanical failure of two synthetic silica glasses: Vycor and Gelsil, and a natural sandstone rock, during soft uniaxial compression at slow driving rates of 0.7 (Gelsil), 2.4 (sandstone) and 5.7 (Vycor) kPa/s. The three samples are predominantly composed of silica ($SiO_2$), presenting different minerology and microporous structure \cite{PRE_jordi, Prl_18}. 

Vycor is a thirsty silica glass synthesized via phase separation of a Na2O-B2O3-SiO2 mixture, quenched and leached with an acid solution leaving a 98\% pure SiO2 skeleton and an average pore diameter of 7.5nm. Sample V32 corresponds to a rectangular prism of base $17.2 mm^2$ and height 5.65 mm was cut from a larger sample. 
Gelsil is another thirsty silica glass fabricated by the hydrolization of silica, followed by condensation and heat treatment, leading to a consolidated assembly of pure silica spheres. The average pore diameter of 2.6 nm \cite{PRE_jordi}, with a total porosity volume of 0.36. Density is measured to be $1.6 g/cm^3$. An original Gelsil monolith was cut in a small rectangular prism of base $46.7mm^2$ and height 6.2 mm (sample Gelsil26). 
Vycor and Gelsil samples were cleaned with a 30\% solution of H2O2, during 24 h and dried at 130◦C before the experiment. 
The sandstone sample (SR2) is a natural rock extracted from a Devonian outcrop in Arran (UK). The main components are 79\% quartz, 5\% feldspar, 11\% clay and cemented by carbonates with a notable presence of iron oxide resulting on a red tonality. The average grain size is 0.3 mm with a pore diameter estimated below 0.1 mm, and a porosity of 0.17. The density is measured to be $2.2 g/cm^3$.

During the experiments, the samples are placed between two aluminum plates where the bottom plate is hanging statically from the load cell (1 kN range and 1N resolution) at the top of the arrangement. The upper plate is pulled downward at constant stress rate which is tracked by the load cell. The sample height (h) is tracked by a laser extensometer (Fiedler Optoelektronik) with a nominal resolution of 100 nm \cite{PhiMag2013, PRE_jordi, Prl_13}.  
The ultrasonic acoustic signal is continuously measured by high frequency piezoelectric transducers acoustically coupled to each aluminum plate. The signals from the transducers are preamplified (60 dB), band filtered (between 100 kHz and 2 MHz) and transferred to a PCI-2 acquisition system from Europhysical Acoustics (Mistras group) working at a time resolution of 40 MHz.

A catalog of individual a.e. events is recorded as a point process by thresholding the acoustic signal $V (t)$. Each event is located in time $t_i$ given the first hitting to the threshold, and marked by the duration $D_i$ and the a.e. energy $E_i$ measured as:
\begin{equation}
E_i \propto \int^{t_i+D_i}_{t_i} |V (t)|^2 dt.
\end{equation}

Previous publications \cite{PRE_nataf, Prl_18} reported (i) the existence of interevent triggering phenomena in the a.e. catalog, i.e., aftershock sequences, (ii) a stationary power-law distribution of E, and (iii) an acceleration of the activity preceding significant strain drops, i.e., foreshock activity. The three observations are consistent with the presence of transient hardening in the mean-field approximation \cite{Prl_18}. 

Given the accelerated foreshock activities, if we look at the (acoustic) energy release rate $dE/dt$, where $E$ is the cumulative energy released during a predefined interval $dt$ (taken here as 50), there will be more number of events within that interval as a main shock is approached. This, in turn, would result in a divergence in the energy release rate. In Fig. \ref{sr2}(a), the time series of energy release rate is shown for the sandstone sample. From the time series, four peaks were identified (as in \cite{PRE_nataf}; peaks appearing very close to one another were discarded). In the subsequent components of the figure (Fig. \ref{sr2}(b-e)), the energy release rate values before the individual peaks were considered and the inequality indices $g$ and $k$ were measured. However, for the crossing of the $g$ and $k$ values to occur, the divergences need to be sufficiently strong (see \cite{PhysRevE.109.044113}), which was not always the case for just the energy release rates. Therefore, the inequality indices were calculated for the square of the energy release rate $(dE/dt)^2$ in all cases.

As can be seen from Fig. \ref{sr2}(b-e), the inequality indices cross before a major event (the last one in the section). The events are colored in green and red, prior to and subsequent to the crossing of $g$ and $k$. Similar analysis were also done for the vycor and gelsil samples (see Supplemental Materials (SM)). The crossing of the inequality indices were seen for each peak of each sample. On average, a warning time window (the time between the crossing and the main shock) of about 5-10 seconds were noted (this also depends on the rate of compression). As mentioned earlier, the samples were of different compositions and the compression rates were also different. Nevertheless, the method is universally applicable.

The underlying failure process of these materials was reproduced in a prototype fiber bundle model with equal load sharing and considering generalized viscoelastic Zener solids instead of elastic fibers \cite{PRE_jordi}. The introduction of viscoelasticity splits large bursts of fibers breaking into smaller sub-avalanche events emitting a.e., trading the characteristic divergence of burst sizes with the increase in activity observed in the experiments \cite{PhiMag2013}. In the similar way as in the cases of the experiments, the energy release rates and the corresponding inequality indices were calculated (see Fig. \ref{model}). The $g$ and $k$ values cross prior to the breakdown of the model and the energy release rate value are depicted in green and red separating the warning period. 

We can quantify the effectiveness of the warning signal (crossing of $g$ and $k$) as a precursor to an imminent major damage to the sample: both digital and actual rocks. To do that, we define a quantity $Q(E)$ that measures the ratio between the cumulative fraction of a.e. events having energy less than or equal to $E$ that are marked as close to failure (red) and the total number of such events. Fig. \ref{Q_measures} shows, for both the experiments and simulations that $Q$ is a monotonically increasing quantity, reaching almost 1 for high values of energy. This shows that the events that resulted in a high value of a.e. emission are the ones mostly classified as too close to the major damage. In other words, the warning method applied here correctly segregates events that cause most of the damages.  

In conclusion, the inequality in the diverging energy release rate during compression failure of nanoporous materials, quantified using social inequality indices (Gini and Kolkata indices), show a precursory warning signal prior to a large event (main shock). The applicability of this method of finding precursory signals is verified in several samples differing in their constituents and compression rates and also using a viscoelastic fiber bundle model simulations. These results suggest that the methods can be applied as a framework for precursory warning signals in a broad class of disordered materials with foreshock activities.

\end{document}